\documentclass{wqcd03}                 
\def\nslash{\rlap{\hspace{0.02cm}/}{n}}
\def\dslash{\rlap{\hspace{0.08cm}/}{D}}
\def\beq{\begin{eqnarray}}
\def\eeq{\end  {eqnarray}}

\def\pr{^{\prime}}
\def\GeV{{\rm GeV} }

\def\non{\nonumber}

\def\lqcd{\Lambda_{\rm QCD}}

\newcommand\npb{Nucl.\ Phys.\ B }
\newcommand\npps{Nucl.\ Phys.\ B (Proc.\ Suppl.) }
\newcommand\plb{Phys.\ Lett.\ B }
\newcommand\zpc{Z.\ Phys.\ C }
\newcommand\prd{Phys.\ Rev.\ D }

\newcommand\prl{Phys.\ Rev.\ Lett. }

\title{The factorization in exclusive B decays: a critical look}

\author{Zheng-Tao Wei }

\address {Departamento de F\'{\i}sica Te\'orica, Universidad
 de Valencia, \\ E-46100, Burjassot, Valencia, Spain}

\begin{document}

\begin{abstract}

I review the theoretical ideas and concepts along the line of
factorization in the exclusive B decays. In order to understand
the naive factorization, the effective field theories and the
perturbative method of QCD are introduced and developed. We focus
our discussions on the large energy effective theory, the QCD
factorization approach and the soft-collinear effective theory.

\end{abstract}

\maketitle

\section{Introduction}

The exploration of CP violation and determination of the
Cabibbo-Kobayashi-Maskawa (CKM) matrix elements motivate extensive
interests of B meson decay. From another point of view, B decays
provide a good place to study the fruitful dynamics of QCD. Up to
now, we have not a truly successful method to calculate the
non-perturbative QCD and the mechanism of quark confinement is
still unknown. The study of exclusive B decays is usually
difficult because of the complicate QCD dynamics. However, the
experiments from the Belle and Babar collaborations have
accumulated and will continue to accumulate a large amount of data
of B decays. The large theoretical uncertainties cannot compete
with the more precise experimental data. We come to one stage that
experiment goes ahead of theory. The theorists in B physics have
to meet great challenge from the experiment.

The problem of exclusive B decays lies in a very large number of
degrees of freedom. The experiment observes the hadron states such
as B meson and pion, kaon etc. In the QCD Lagrangian, only quark
and gluon degrees of freedom appear. We don't know accurately how
the hadron are formed by quarks and gluons. From the energy scale
standpoint, the B decays usually contain many scales: the weak
interaction scale $m_W$, the $b$ quark mass $m_b$, the QCD scale
$\lqcd$ and possible intermediate scales due to the soft spectator
quark in B meson. The momenta of quarks or gluons are not
restricted. They can be highly virtual, very soft or highly
energetic but collinear to the fast moving pion. The fact that we
have to treat all the degrees of freedom in one process if we
think QCD is the correct theory of strong interaction leads to
great theoretical complications.

One method to treat the multi-scales problem is factorization. The
factorization is a key ingredient of perturbative QCD (pQCD)
\cite{CSS,BL}. Its basic idea is to separate the short-distant
dynamics from the long-distance physics. It has been widely used
in the hard QCD processes where the large momentum transfer $Q\gg
\lqcd$ is involved. Another method is the effective field theory.
It is a useful toll to study the process with several separate
scales. The heavy quark effective theory (HQET) \cite{Georgi} is a
low energy effective theory. It allows model-independent
predictions in some cases of the heavy meson system, such as $B\to
D$ form factor at zero recoil. The developments of the two methods
are nearly independent although some ideas in them are related. As
we will show that these two lines of thought converge in the study
of exclusive B decays,.

The factorization had been introduced in exclusive B decays for a
long time. The old form which we call the naive factorization
approach is to divide a hadronic matrix element into the
multiplication of a form factor and hadron decay constant. This
idea influences B physics for more than 30 years. Much efforts
were done to interpret and generalize it. Now, the idea of
factorization has been developed as a central idea of B physics.
In this talk, we will discuss the theoretical struggle of studying
the exclusive B decays along the line of factorization. The
success and limitations of each theoretical approach will be
analyzed. We are focus on the conceptual developments from the
naive factorization approach to the QCD factorization approach. We
will show how the effective field theory enters into B physics and
modifies our view.

\section{The naive factorization approach and the large energy
 effective theory}

The first thing to do in B decays is to integrate out the heavy
degrees of freedom of W, Z bosons and top quark in the standard
model. The method is to construct an effective theory where the
above heavy particles do not appear. The theoretical technic is
mature now. It uses the operator product expansion (OPE) and
renormalization group equation (RGE). For non-leptonic B decays,
the
relevant effective weak Hamiltonian is %
\beq %
{\cal H}_{eff}=\frac{G_F}{\sqrt 2}\sum_i V^i_{\rm CKM}C_i(\mu)Q_i.
\eeq %
where $G_F$ is the Fermi constant and $Q_i$ are current-current
operators. The scale $\mu$ is chosen of order of $m_b$. The
amplitude of $B\to M_1M_2$ decay is%
\beq %
A(B\to M_1M_2)=\frac{G_F}{\sqrt 2}\sum_i V^i_{\rm CKM} \non \\
  \times C_i(\mu)\langle M_1M_2|Q_i|B\rangle(\mu).
\eeq %
where $\langle M_1M_2|Q_i|B\rangle$ are hadronic matrix elements.
The remained work is to calculate the hadronic matrix element.

\subsection{The naive factorization approach}

The introduction of factorization to simplify the the hadronic
matrix element may be firstly given in \cite{HS} up to knowledge
of the author. I cannot trace out this history but refer to
\cite{BSW} as our start of discussion. Bauer, Stech and Wirbel
consider the non-leptonic two meson decays where the final mesons
are energetic. They made assumptions that only the asymptotic part
of the hadron field is effective and the current are proportional
the hadron field. All the initial state interaction and final
state interactions are neglected. Based on the above assumptions,
one hadron and its associated current are separated out. The
hadronic matrix element is factorized into a multiplication of
decay constant and the form factor which represented by the matrix
element of the other current. Take $\bar B^0\to \pi^+\pi^-$ decay
as an example, %
\beq %
\langle \pi^+\pi^-|(\bar u b)_{\rm V-A}(\bar d u)_{\rm V-A}|
  \bar B^0 \rangle = ~~~~~~~~\non \\ ~~~~~~~~
 \langle \pi^-|(\bar d u)_{\rm V-A}|0\rangle
 \langle \pi^+|(\bar u b)_{\rm V-A}|\bar B^0\rangle.
\eeq %
The idea of the above factorization is simple but it has a deep
influence. The application of the above naive factorization
approach into the non-leptonic two body B decays is successful in
early days of B physics when the experimental data are rare. For a
long time, this approach is nearly the only method to give a
theoretical prediction of exclusive non-leptonic B decays although
the accuracy is at the qualitative level for many processes.

The factorization approach plays a similar role as the Feynman's
parton model in DIS, we can call this naive approach as the parton
model in B physics. One might expect that the factorization is a
limit case of a more general theory. The understanding of the
factorization from field theory of QCD is a long way. The first
step comes from Bjorken's intuitive space-time picture
\cite{Bjorken}. It is Bjorken who proposed the famous scaling in
DIS which lead to the rise of QCD. For $\bar B^0\to \pi^+\pi^-$
decay, the quark level decay is $b\to u+\bar u d$.  In order to
form the final energetic hadron, the quark pair $\bar u d$ has to
choose a nearly collinear configuration. Because the pions move
fast, the formation time of $\pi^-$ will be long because of the
relativistic time-dilation. The dilation ratio is
$m_b/\lqcd\approx 20$. That means the hadronization occurs $20$ fm
away from the remained system. Before the hadronization, the $\bar
u d$ quark pair produced from the pointlike, color-singlet weak
interaction is a a small color dipole. The small color dipole has
little interaction with the other quarks. The above consideration
is usually called ``color transparency argument". From the above
argument, one may guess that the factorization approach is the
leading order contribution of heavy quark limit where $m_b\gg
\lqcd$ and the non-factorizable corrections come from the
interactions of small color dipole with the remained quarks at
short distance.

What is color transparency? It is a concept outside of B physics.
According to the discussions in \cite{BBGG}, the color
transparency is a phenomenon of pQCD \cite{BL}. It says that a
small color-singlet object can pass freely through nucleon target
as if the target is transparent. The large target acts as a filter
which removes the large transverse separation component of the
hadron. The $\bar B^0\to D^+\pi^-$ decay provides a similar
environment. $B\to D$ transition is at long distance. The
energetic $\pi^-$ selects the $\bar u d$ quark pair at small
transverse separations. The long distance processes caused by
emitting or absorbing soft gluons have destructive effects and
cancel.

In \cite{PW}, Politzer and Wise apply  the pQCD method into the
exclusive processes in $\bar B^0\to D^+\pi^-$ decay. They proposed
a factorization formula that the hadronic matrix element can be
written as the product of a matrix element of $B$ and $D$ mesons
in HQET and a convolution by a hard scattering amplitude T and the
pion distribution amplitude $\phi_{\pi}(x)$ as %
\beq \label{eq:PW}%
\langle D^+\pi^-|(\bar c b)_{\rm V-A}(\bar d u)_{\rm V-A}|
   \bar B^0 \rangle=
\langle D^+|(\bar c_v b_{v'})_{\rm V-A}|\bar B^0\rangle \non \\
 \times \int_0^1 dx~ T(x, m_c/m_b, \mu)\phi_{\pi}(x, \mu).s
\eeq %
where $c_v, b_{v'}$ are effective fields for heavy quarks in HQET.
They point out the above factorization formulae is the leading
order result in $\lqcd/m_b$. One loop calculate is done and the
result show $\alpha_s$ correction to leading contribution is
small. However, they don't give proof of the factorization.

\subsection{The large energy effective theory}

The success of HQET motivates theorists to use effective field
theory into wider range of application. Dugan and Grinstein had a
new idea to establish a foundation for factorization on the
effective field theory. They use the effective field theory to
replace the intuitive ``color transparency argument". In
\cite{DG}, Dugan and Grinstein proposed a large energy effective
theory (LEET) to describe the interaction of the energetic
collinear quark with soft gluons. They consider one kinematic case
that the energy of the collinear quark is much lager than the
momentum of soft gluon, i.e., $E\gg \lqcd$. The central idea in
LEET is that the energy of the collinear quark is unchanged which
is analogous to the velocity superselection rule
in HQET. The LEET is very similar to the HQET. The LEET Lagrangian is %
\beq \label{eq:LEET}%
{\cal L}_{LEET}=\psi^{\dagger}in\cdot D\psi.
\eeq %
where $n$ is a light-like vector which the direction is along the
motion of the collinear quark and the collinear field $\psi$
satisfies $\nslash \psi=0$.

If we choose the light-cone gauge $n\cdot A=0$, the soft gluons
decouple from the collinear and factorization is a trivial result.
From this point of view, the color transparency is explained by
that only the longitudinal gluons couple to the collinear quark
and thus decouple. Although the proof of factorization in this way
is too simple to be correct, the LEET is very impressive. It
provide a new view of factorization. In pQCD, the diagrammatic
analysis \cite{CSS} is the most familiar method to prove
factorization. The LEET provide an operator description and the
result is automatical to all orders. The proposal of an effective
Lagrangian permits us to use gauge symmetry at the Lagrangian
level. The proof of factorization can be easily done in a gauge
invariant way, i.e., we need not have to choose the light-cone
gauge $n\cdot A=0$. As we will show, the LEET is one part of the
soft-collinear effective theory. The LEET Lagrangian given in Eq.
(\ref{eq:LEET}) is just the leading order result.

The biggest problem of the LEET is that it cannot reproduce the
long-distance physics of QCD. The reason is simple because it
misses the collinear gluon degrees of freedom. Without collinear
gluon, the collinear quark pair $\bar u d$ can not form the
energetic bound state $\pi^-$. The neglect of collinear gluon is
pointed out in \cite{AC}. Aglietti and Corb\'{o} point out one
problem of LEET and modify the LEET by including the transverse
degrees of freedom \cite{AC}. The improved LEET Lagrangian is
given by %
\beq \label{eq:LEET2}%
{\cal L}=\psi^{\dagger}\left(in\cdot D
  +\frac{D_T^2}{2E}\right)\psi.
\eeq %
where $E$ is the energy of the collinear quark. In this modified
version, the collinear gluon is still missing. The effective
theory which includes the collinear gluon is more complicated than
Eq. (\ref{eq:LEET2}) because the energy $E$ is not conserved.
Except the missing of the collinear gluon degrees of freedom, both
the LEET and its modified version have the problem that there is
no systematic power counting to support them. They have to wait
for the next step development.

\section{The QCD factorization approach and the soft-collinear
  effective theory}

The application of the pQCD method in \cite{BL} into the exclusive
B decays had been explored by many theorists. There are several
different perturbative approaches appeared in the literatures. Due
to the scope of this review, we focus our discussions on the
recently proposed QCD factorization approach in
\cite{BBNS1,BBNS2}.

\subsection{The QCD factorization approach}

Beneke, Buchalla, Neubert and Sachrajda want to establish a
rigorous framework for the exclusive non-leptonic B decays. The
basic idea of the QCD factorization approach is that in the heavy
quark limit the naive factorization is the lowest order
approximation and the corrections to the naive factorization can
be formulated as a factorization form up to corrections of order
$\lqcd/m_b$. The heavy quark limit $m_b\gg \lqcd$ is the kinematic
foundation of the QCD factorization approach. For $\bar B^0\to
\pi\pi$ decay,
the factorization formula is %
\beq \label{eq:QFA}%
\langle \pi\pi|Q_i|\bar B^0\rangle=F^{B\pi}(0)\int_0^1 dx~T^I_i(x)
 \phi_{\pi}(x)~~~~~~~~~~\non \\
 +\int_0^1 d\xi dxdy~T^{II}_i(\xi,x,y)
 \phi_B(\xi)\phi_{\pi}(x)\phi_{\pi}(y).
\eeq %
where $F^{B\pi}$ is a $B\to \pi$ form factor at $q^2=0$,
$\phi_{\pi(B)}$ are light-cone distribution amplitudes of the pion
and B meson. The $T_i^{I(II)}$ are perturbatively calculable hard
scattering kernels. Compared to Eq. (\ref{eq:PW}), the difference
lies in the second term due to the hard spectator correction.

Ref. \cite{BBNS2} can be considered as a systematic introduction
of the pQCD method into B physics for the first time. In
\cite{BBNS2}, a power counting is used to argue the validity of
the QCD factorization approach. This power counting largely uses
the endpoint behavior of the distribution amplitudes of mesons.
Annihilation diagrams and higher Fock states of the mesons are
proved to be power suppressed. The hard spectator interaction in
$\bar B^0\to D^+\pi^-$ decay is suppressed if one assumes $c$
quark is heavy.

The validity of factorization can not be based on the intuitive
arguments only. It should be proved to all orders that all the
soft and collinear divergences cancel or can be absorbed into the
universal non-perturbative functions and the hard scattering
kernels are infrared insensitive. For $B\to \pi\pi$ decays, the
factorization is proved to hold in $\alpha_s$ order. But it is not
sufficient to guarantee the validity of factorization because all
the soft and collinear divergences cancel is not general. The
general case is that the infrared divergences may not cancel but
they can be separated out and absorbed into the definition of the
non-perturbative functions. Up to now, the factorization beyond
$\alpha_s$ order in $B\to \pi\pi$ decays has not been truly
proved.

In \cite{BBNS2}, the factorization proof for $B\to D\pi$ decays at
two-loop order is given. Two-loop order is equivalent to
$\alpha_s^2$ order because the hard spectator interaction is power
suppressed for the heavy-light final states. The authors consider
62 ``non-factorizable" diagrams at two loop order. Maybe their two
loop order proof of factorization is the most detailed analysis in
the literatures up to my knowledge. The eikonal approximation and
the Ward identity are used implicitly. At two-loop order, the
infrared divergences in the soft-soft, soft-collinear and
collinear-collinear momentum regions cancel. The hard-collinear
and hard-soft contributions contain non-cancelling infrared
divergences. They can be factorizaed out and absorbed into the
definition of distribution amplitude and form factor respectively.
The proof of factorization in the diagrammatic analysis is
intuitive, but it is impossible to go to all orders. A
factorization proof of all orders is necessary.

\subsection{The soft-collinear effective theory}

From the LEET, we know that the effective field theory can
simplify the analysis of the infrared physics and the
factorization in it is automatic to all orders. One natural idea
is: can we construct an effective field theory for the soft and
collinear particles which reproduce all the infrared physics of
QCD and can simplify the factorization proof?

Bauer el al. aim at developing a soft-collinear effective theory
by generalizing the idea of LEET. They start from the study of
summing Sudakov double logarithms in inclusive $B\to X_s\gamma$
decay. For $B\to X_s\gamma$ decay near the endpoint of the photon
spectrum, it contains energetic light particle. The Sudakov double
logarithms will appear in one loop order due to the
non-cancellation of the soft and collinear divergences. The large
double logarithms make the perturbation expansion ill-behaved and
need to be resumed to all orders. The Sudakov resummation is
usually considered as an important but difficult part in the
conventional pQCD method. By matching the full theory to a new
effective theory, the large logarithms cancel and the Sudakov
doule logarithms are summed by using the renormalization group
equations \cite{B1}. One important thing is that  Sudakov
resummation in the effective theory is simpler than that in the
full theory.

The analysis in \cite{B1} also shows another important thing that
the effective field theory can be used in the case where the
Wilson's short-distance operator product expansion (OPE) is not
applicable. The idea is matching the full theory onto the
effective theory where the effective operators provide the
long-distance information of QCD. To this externt, the effective
field theory develops the idea of OPE.

The formalism of the soft-collinear effective theory is provide in
\cite{B2}. At this stage, only the ultrasoft gluons are included.
The lowest order effective Lagrangian is written by %
\beq %
{\cal L}_{eff}=\bar \xi_{n_-,p\pr} \left [ ~in_-\cdot D
  +i\dslash_{c\bot}\frac{1}{~in_+\cdot D_c}~ i\dslash_{c\bot}
  \right ]  \\
  \times\frac{\nslash_+}{2}~\xi_{n_-,p}~.\non
\eeq %
where $in_-\cdot D=in_-\cdot \partial+gn_-\cdot(A_c+A_{us})$,
$in_+\cdot D_c=\bar{\cal P}+gn_+\cdot A_c$, $i\dslash_{c\bot}=\bar
{\cal P}_{\bot}+gA_{c\bot}$ and $n_-, n_+$ are two light-like
vectors. Because the formulation is given in a hybrid
position-momentum space. A momentum label operator ${\cal P}$ has
to be introduced because the large momentum is not conserved.
Beneke, Chapovsky, Diehl and Feldmann developed a position space
formalism in order to avoid the complicate momentum label operator
in \cite{BCDF}.

In \cite{Wei}, I propose a soft-collinear effective theory which
includes the soft gluons in the position space. The final SCET
Languagian is%
\beq \label{eq:SCET}%
{\cal L}_{SCET}=\bar{\xi}\left [ ~in_-\cdot D
  +i\dslash_{\bot}\frac{1}{~in_+\cdot D}~ i\dslash_{\bot}
  \right ]\frac{\nslash_+}{2}~\xi.
\eeq %
where the covariant derivative is defined by
$D=\partial-igA_c-igA_{(u)s}$.

From the above effective Lagrangian, it is easy to obtain the
(ultra)soft and collinear Wilson lines. The SCET has rich symmetry
structures. The Lorentz and gauge invariance are interesting in
SCET. There is a new symmetry, scale symmetry. This symmetry
provide an interpretation of the Bjorken's scaling and the scaling
law in high energy scattering.

The application of SCET to prove the factorization in $B\to D\pi$
decays is given in \cite{B3,B4}. I give a factorization proof in
DIS in \cite{Wei} and SCET can also be applied into multi-body B
decays, such as $B\to DKK$ decays \cite{DKK}.

\section{The questions about the QCD factorization approach}

Although SCET provides a new theoretical framework, the practical
calculations of exclusive B decays still rely on some
factorization formulae, such as the QCD factorization approach. By
use of this opportunity, I want to express my personal opinions on
the QCD factorization approach. I will ask some conceptual
questions about it.

I. Is the factorization in $B\to \pi\pi$ decays proved?

~~~~As we have discussed above, the factorization in $B\to \pi\pi$
decays is given only at $\alpha_s$ order. This is not sufficient
for the validity of factorization. We can say that the QCD
factorization approach has not been proved as a ``theory''. Even
for $B\to D\pi$ decays, the factorization are based on some
kinematical assumptions, such as $m_b\sim m_c\to \infty$ or the
ratio $m_b/m_c$ is fixed. The real world is: $m_b\approx 4.5\GeV$,
$m_c\approx 1.5\GeV \sim\sqrt{m_b\lqcd}$. Although one can assume
a limit case for the theoretical purpose, the relation between the
ideal world and the real world needs to be explored.

II. Is the QCD factorization approach correct in the $m_b\to
 \infty$ limit?

~~~~This question is related to the first question. This time we
will not concern the technical complications about the
factorization proof but the general argument of factorization. The
argument is  Bjorken's ``color transparency". One strange thing
for me is why the color transparency can be applied for one pion
but not for another pion in $B\to \pi\pi$ decays. The two pions
are both energetic and have the same momentum in the rest frame of
B meson, but they are treated unequally. The factorization formula
in Eq. (\ref{eq:QFA}) contains two terms, one is related to the
naive factorization approach, the other likes the hard scattering
approach. The two different terms make the factorization formula
un-natural. One associated problem is: the renormalization scale
$\mu$ in first term is at order of $m_b$, while in the next term
$\mu\sim \sqrt{m_b\lqcd}$. There are two large scales in
$B\to\pi\pi$ decays. So it is a multi-scales problem. My oppinion
is that we should be more serious about his problem.

III. Is $B\to\pi$ form factor baisc?

This is a very controversial topic. In the QCD factorization
approach, it is assumed as a basic function. But this function is
different from other fundamental non-perturbative QCD fucntions
such as the Isgur-Wise function, pion distribution amplitudes etc.
$B\to\pi$ form factor is not and can not be represented as a
dimensionless function. From the soft-collinear effective theory
which is scale invariant, the basic non-perturbative functions
should be dimensionaless except the dependence on the
renormalization $\mu$. The $B\to\pi$ form factor does not satisfy
this criterion. In fact, $B\to\pi$ form factor contains very
complicate QCD dynamics which includes scales of $m_b,
\sqrt{m_b\lqcd}$ and scales between $\sqrt{m_b\lqcd}$ and $\lqcd$.
From this point of view, the QCD factorization seems more like a
phenomenological ``approach" rather than a ``theory".

\section{Conclusions}

We have reviewed the theoretical developments from the naive
factorization approach to the soft-collinear effective theory. In
order to understand the factorization in exclusive B decays, new
ideas and new theories or approaches are produced. The
soft-collinear effective theory is a rigorous theory. It provides
a theoretical foundation of the factorization theorem in pQCD and
a unified framework to study the inclusive and exclusive hard QCD
processes. The soft-collinear effective theory is the second
theoretical contribution of B physics to QCD. The first is HQET.

The QCD factorization approach can not solve the full
complications of exclusive B decays. The application of the
soft-collinear effective in B decays is still limited because the
momentum of quarks in hadrons depend on the non-perturbative
dynamics for exclusive processes. The final solution of B decays
must rely on the solution on the quark confinement and
non-perturbative problems.  If these fundamental problems of the
strong interaction were solved by the theorists in B physics, it
will not be a miracle. This is the great challenge of B physics.

\section*{Acknowledgments}

The author acknowledges a postdoc fellowship of the Spanish
Ministry of Education. This research is supported by Grant
FPA/2002-0612 of the Ministry of Science and Technology.

\end{document}